\documentclass[preprints,article,accept,moreauthors,pdftex,10pt,a4paper]{Definitions/mdpi} 
\firstpage{1} 
\pubvolume{xx}
\issuenum{tba}
\articlenumber{tba}
\pubyear{2019}
\copyrightyear{2019}
\history{Received: tba; Accepted: tba; Published: tba}

\usepackage[utf8]{inputenc}
\usepackage{textcomp}
\usepackage{textgreek}
\usepackage{amsmath,bm}
\usepackage{amssymb}
\usepackage{graphicx}
\usepackage{mathabx}  
\usepackage{float}
\usepackage[export]{adjustbox}
\usepackage{color}
\usepackage{makeidx}
\usepackage[caption=false]{subfig}

\usepackage{nomencl}

\renewcommand\nomgroup[1]{%
  \ifthenelse{\equal{#1}{A}}{%
    \item[\textbf{Acronyms}]}{
  \ifthenelse{\equal{#1}{D}}{%
    \item[\textbf{Dimensionless Quantities}]}{
  \ifthenelse{\equal{#1}{R}}{%
    \item[\textbf{Roman Symbols}]}{
  \ifthenelse{\equal{#1}{G}}{%
    \item[\textbf{Greek Symbols}]}{
  \ifthenelse{\equal{#1}{S}}{%
    \item[\textbf{Superscripts}]}{
  \ifthenelse{\equal{#1}{U}}{%
    \item[\textbf{Subscripts}]}{
  \ifthenelse{\equal{#1}{X}}{%
    \item[\textbf{Other Symbols}]}{
  {}}}}}}}}}

\makeindex
\makenomenclature

\Title{Modeling the excess velocity of low-viscous Taylor droplets in square microchannels}

\Author{Thorben Helmers$^{1}$\orcidA{}*, Philip Kemper $^{2}$, Jorg Th\"oming $^{2}$\orcidC{} and Ulrich Mie\ss ner $^{1}$\orcidD{}}

\AuthorNames{Thorben Helmers, Philip Kemper, Jorg Thöming and Ulrich Mießner}

\address{%
$^{1}$ \quad Department of Environmental Process Engineering, University Bremen, Leobener Str. 6, 28359 Bremen, Germany, sekr@uvt.uni-bremen.de\\
$^{2}$ \quad Department of Chemical Engineering, University Bremen, Leobener Str. 6, 28359 Bremen, Germany, uft@uni-bremen.de}
\corres{Correspondence: helmers@uvt.uni-bremen.de; Tel.: +49-421-218-63337}

\abstract{Microscopic multiphase flows have gained broad interest due to their capability to transfer processes into new operational windows and achieving significant process intensification. However, the hydrodynamic behavior of Taylor droplets is not yet entirely understood. In this work, we introduce a model to determine the excess velocity of Taylor droplets in square microchannels. This velocity difference between the droplet and the total superficial velocity of the flow has a direct influence on the droplet residence time and is linked to the pressure drop. Since the droplet does not occupy the entire channel cross section, it enables the continuous phase to bypass the droplet through the corners. A consideration of the continuity equation generally relates the excess velocity to the mean flow velocity. We base the quantification of the bypass flow on a correlation for the droplet cap deformation from its static shape. The cap deformation reveals the forces of the flowing liquids exerted onto the interface and allows estimating the local driving pressure gradient for the bypass flow. The characterizing parameters are identified as the bypass length, the wall film thickness, the viscosity ratio between both phases and the $Ca$-number. The proposed model is adapted with a stochastic, metaheuristic optimization approach based on high-speed camera measurements. In addition, our model is successfully verified with published empirical data.}

\keyword{Taylor flow; droplet excess velocity; droplet velocity model; microfluidics}

\begin{document}


\section{Introduction}
Microscopic multiphase flows facilitate a wide field of possible applications since they provide short diffusion layers within the flow structures. This enables high mass and heat transfer rates \citep{Haase.2016,SattariNajafabadi.2018} for several applications ranging from extraction \citep{Kralj.2007} and multiphase catalyst reactions \citep{Kobayashi.2006} to improved unit operations like mixing tasks \citep{WONG.2004}. The distinct features allow performing reactions at new process windows with fewer hazards or higher selectivity \citep{Sun.2016}. The specific flow conditions can furthermore serve for cell isolation \citep{Chen.2014}, genetic analysis \citep{Hosokawa.2017} and reaction screening in a droplet chain \citep{Jin.2018}. In contrast to large scale multiphase flows, microscopic flows are much easier to predict as there are no complex interactions such as swarm turbulence \citep{Kuck.2018,Miener.2017} commonly found in bubble columns. In fact, the reproducibility of e.g. Taylor flows is a key for the application of microscopic multiphase flows \citep{Kreutzer.2005b}.

In the Taylor flow regime, the disperse phase is separated from the wall by a thin wall film and does not fill the entire cross-section of the microchannel. The remaining space between the droplet and the microchannel corners is occupied by the continuous phase, which are referred to as gutters \citep{VANSTEIJN.2008}. The droplets are typically longer than the channel diameter, which leads to separated elongated disperse phase instances. The continuous phase segments between the droplets are called slugs. Taylor flows are mainly established in circular capillaries or rectangular microchannels whereas especially the hydrodynamics of moving Taylor droplets in circular capillaries and the role of the thin wall film have been intensively studied \citep{Abiev.2017}.

Chemical reactions on the microscale are often performed in monolith reactors as a catalyst support. Within these reactors, a high number of parallelized channels with hexagonal or square channel cross-section offer a high specific reaction area for wall placed catalysts at small wall thickness. This results in a better heat transfer through the walls and better mechanical stability than circular capillaries \citep{Gascon.2015}. For process control and stabilization, as well as precise reactor design, knowledge of the underlying fluidic terms is crucial. The high grade of parallelity complicates prediction of the hydrodynamics and the resulting pressure drop \citep{Kreutzer.2005a,Huerre.2014}.

Besides disperse phase size distribution and formation frequency \citep{Fu.2015}, the actual droplet velocity is essential for the droplet residence time in the reactor. It determines the contact time of the educts and influences the pressure drop of the reactor \citep{adosz.2018}. Within a parallel reactor, an exact knowledge of the pressure drop is especially necessary since a steady educt supply for each individual single reactor is needed to ensure stable and efficient working conditions \citep{Schubert.2016}.

Several publications have dealt with the droplet velocity in rectangular capillaries and observed a droplet velocity mostly faster than the superficial velocity (see Sec. \ref{sec:concept_exc}). For flows within circular capillaries, where only a thin wall film is present, this velocity difference is well understood \citep{Abiev.2017b}, while for rectangular microchannel a variety of explanations exist, that mostly correlate the relations from measurements \citep{Jose.2014}. This complicates the transfer of results to other flow applications or altered process parameters since local and instantaneous hydrodynamic parameters are mostly not taken into account by the models and correlations.

This work aims to establish a model to determine the droplet velocity from the actual flow conditions: e.g. droplet length, material properties, and the $Ca$-number. In a first step, we develop a concept for the relative droplet velocity, which bases this velocity on extrinsic parameters, allowing symmetrical scaling. From this concept, we identify the bypass flow through the gutters as well as the film-thickness as the prominent parameters for the excess velocity.

In the next step, we develop a model that uses the local surface curvature of the gutters to retrieve the local pressure at the entrance and outlet of the gutter's corner as a driving force. The bypassing gutter flow is calculated based on the counterplay of this driving force, the gutter length and a viscosity correlated resistance factor $\beta$. The local droplet curvature at the gutter entrances is derived with an analytical interface shape approximation \citep{Miessner.2019} and a correlation for the droplet cap curvature based on the $Ca$-number from our previous work \citep{Helmers.2019}. The model is successfully validated by high-speed camera measurements.


\section{Hydrodynamic fundamentals of Taylor Flows}\label{sec:fluid_bas}
The Taylor flow regime in rectangular micro-channels is mainly influenced by surface tension forces rather than inertia forces. In the Taylor flow regime, a droplet fills nearly the whole cross-section of a hydrodynamic channel, while the continuous phase occupies the gutters and a thin wall film. The droplets are divided by the slugs, consisting only of liquid from the continuous phase.

The interaction between interfacial and viscous forces is described by the Capillary number $Ca$.

\begin{equation}\label{eq:1}
Ca=\frac{u_{0}\eta_{c}}{\sigma}
\end{equation}
\par

\nomenclature[D]{$Ca$}{Capillary number [$-$]}
\nomenclature[R]{$u$}{velocity [$mm \cdot s^{-1}$]}
\nomenclature[U]{$0$}{superficial/total}
\nomenclature[G]{$\eta$}{dynamic viscosity [$Pa \cdot s$]}
\nomenclature[G]{$\sigma$}{interfacial tension [$N \cdot m^{-1}$]}

Herein, $\eta_{c}$ represents the dynamic viscosity of the continuous phase, $\sigma$ the interfacial tension between both phases and $u_{0}$ the total superficial flow velocity: 

\begin{equation}\label{eq:sup_vel}
u_{0}=\frac{Q_{0}}{A_{ch}}=\frac{Q_{d}+Q_{c}}{W \cdot H}= \frac{Q_{d}+Q_{c}}{H^2 \cdot ar}
\end{equation}
\par

\nomenclature[R]{$Q$}{volume flow rate [$\mu l \cdot min^{-1}$]}
\nomenclature[U]{$d$}{disperse/droplet phase}
\nomenclature[U]{$c$}{continuous phase}
\nomenclature[R]{$A$}{cross sectional area [$m^{2}$]}
\nomenclature[R]{$W$}{channel width [$\mu m$]}
\nomenclature[R]{$H$}{channel height [$\mu m$]}
\nomenclature[U]{$ch$}{channel}
\nomenclature[R]{$ar$}{micochannel aspect ratio}

The superficial velocity $u_0$ is based on the volume flow of the disperse ($Q_{d}$) and continuous ($Q_{c}$) phase as well as the microchannel's cross-sectional area $A_{ch}$ that calculates from the channel width $W$ and channel height $H$ or respectively the channel aspect ratio $ar= \frac{W}{H}$. 
 
 \nomenclature[R]{$ar$}{aspect ratio $[-]$}
 \nomenclature[R]{$ar$}{aspect ratio $[-]$}

For energetic considerations, knowledge of the ratio between inertia and viscous forces is of importance. The Reynolds-number  

\begin{equation}\label{eq:re}
Re=\frac{u_{0} d \rho_c}{\eta_c}
\end{equation}
\par

\nomenclature[D]{$Re$}{Reynolds number [$-$]}
\nomenclature[G]{$\rho$}{density [$kg \cdot m^{-3}$]}
\nomenclature[R]{$d$}{characteristic length [$m$]}

is used to describe the ratio between inertia and viscous forces of the continuous phase with $\rho_c$ being the continuous phase density and $\eta_c$ the continuous phase viscosity. Additionally, the viscosity ratio 

\begin{equation}\label{eq:lambda}
\lambda = \frac{\eta_d}{\eta_c}
\end{equation}
\par

\nomenclature[D]{$\lambda$}{Viscosity ratio [$-$]}

between both phases has a significant influence on the pressure drop and the velocity of a droplet \citep{Direito.2017,Shams.2018,adosz.2018} since it indicates the momentum coupling into the disperse phase. Please note, that the definition of $\lambda$ differs within the mentioned publications.

Considering the overall classification of the applied flow system, the material properties of both fluid phases are of importance. The Ohnesorge-number $Oh$ describes the most prominent material properties for droplets being formed or dispersed \citep{Ohnesorge.1936}. It characterizes the fluidic system independent of current flow or forces and is mostly used when working with surfactants to manipulate the flow properties. Here, we use the $Oh$-number to characterize the continuous phase. 

\begin{equation}\label{eq:Oh}
Oh = \frac{\eta_{c}}{\sqrt{d \rho_{c} \sigma}} = \sqrt{\frac{Ca}{Re}}
\end{equation}
\par

\nomenclature[D]{$Oh$}{Ohnesorge number [$-$]}

In many applications, Taylor droplets in rectangular microchannels move with a velocity different from the superficial velocity, because the droplet does not fill the entire channel cross section and continuous phase can bypass the droplet through the gutters. An early description is given by \citet{Wong.1995}, who dewscribes these phenomena analytically and declare two possible regimes. 

In the first regime, the fluid in the gutters moves slower than the droplet, dissipating kinetic energy. For the gutters, \citet{Abiev.2017b} reported an inversed pressure gradient, indicated by the local surface curvature. For circular microchannels, this results in an inverse flow of the wall film and the droplet moves faster than the superficial velocity.

In the second flow regime, which holds true for long and highly viscous droplets \citep{Wong.1995,Shams.2018,Rao.2018}, more energy is dissipated through the larger wall film area and through viscous dissipation within the droplets. Consequently, the flow in the gutters moves from the droplet back to the front. Thus, the droplet moves slower than the superficial velocity. 

For both regimes, the thin wall film resists the motion of the droplet, resulting in a difference in pressure with higher value at the back and a lower value at the front of the droplet.
The transition between both regimes is described by a critical flow rate and depends on the droplet length and the channel aspect ratio \citep{Wong.1995}.

\citet{Jakiela.2011} focus experimentally on the influence of the momentum coupling between both phases represented by $\lambda$ and also reveal a dependence of the droplet velocity on the droplet length. For short low viscous droplets ($\lambda<1$ ), the droplets move faster than the superficial velocity and the droplet behavior is assigned to the first flow regime. Highly viscous droplets ($\lambda>1$ ) move either faster or slower than the superficial velocity, depending strongly on the droplet length.

\section{Concept of Excess Velocity}\label{sec:concept_exc}

Based on the instantaneous droplet velocity $u_{d}$, that is evident and directly measurable via optical or electrical measurement techniques \citep{Kalantarifard.2018}, different explanatory approaches for the deviation of superficial velocity $u_{0}$ and droplet velocity $u_{d}$ have been reported. 

\citet{Liu.2005} define this relative difference as slipping velocity $u_{slip}$, \citet{Howard.2013} as relative drift velocity $u_{drift}$, \citet{Angeli.2008} as relative bubble velocity $u_{rel}$ and \citet{Abadie.2012} as dimensionless droplet velocity. \citet{Jakiela.2011} focus directly on the ratio of $m = \frac{u_d}{u_0}$ and name this quotient droplet mobility according to \citet{Bretherton.1961}. For this work, we like to summarize these approaches as a slipping velocity:

\begin{equation}\label{eq:uslip}
u_{slip}=\frac{u_{d}-u_{0}}{u_{d}}=1-\frac{u_{0}}{u_{d}}
\end{equation}
\par

\nomenclature[U]{$slip$}{slipping scale concept}

\begin{figure}[htb]
\centering
 \includegraphics[width=0.5\linewidth]{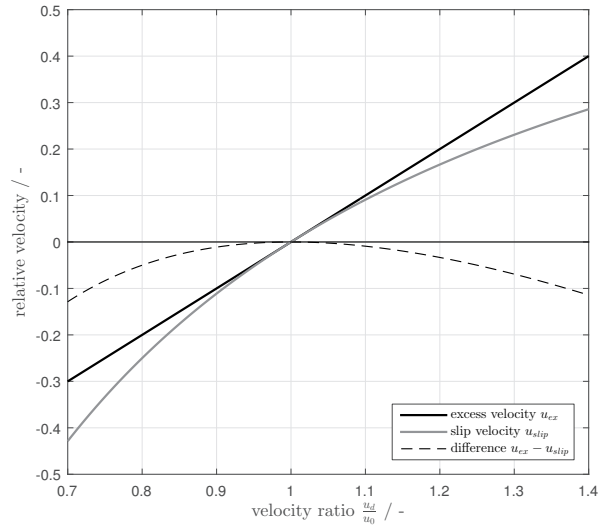}
\caption{Comparison of different concepts of calculating the relative droplet velocity: excess velocity from this work (black solid line), slip velocity (grey solid line) and difference of both concepts (black dashed line)}
\label{fig:comp}       
\end{figure}

In those concepts, the desired quantity is scaled with an intrinsic value such as the instantaneous droplet velocity, which leads to normalization effects  as values $u_{slip}<1$ and $u_{slip}>1$ are not normalized symmetrically (Fig. \ref{fig:comp}). This behavior has to be taken into account when experimental or simulative data is interpreted.

In our approach, we scale the velocity difference with the superficial velocity as an extrinsic property and, staying in the term of extrinsic denomination, define it as an excess velocity $u_{ex}$

\begin{equation}\label{eq:ex}
u_{ex}=\frac{u_{d}-u_{0}}{u_{0}}=\frac{u_{d}}{u_{0}}-1
\end{equation}
\par

\nomenclature[U]{$ex$}{excess scale concept}

In this manner, $u_{ex}$ results values around $0$ for droplet velocities equal the superficial velocity (plug flow), while positive and negative values indicate droplets, which are moving respectively faster or slower than the superficial velocity.

The advantage of this extrinsic concept stands out in a comparison of both approaches (Fig \ref{fig:comp}). The first shown intrinsic concept (slip velocity) leads to an asymmetrical scaling behavior especially for droplets with $u_{d} < u_{0}$, since $u_{slip}<0$ decreases stronger than $u_{slip}>0$ would increase. For applications like balancing or process modeling, a linear behavior is mandatory, to prevent an additional bias of the modeled quantities towards any direction $u_d < u_0 \lor  u_d > u_0$.

A description for the excess velocity can be derived from the volume flows around moving droplets (Eq. \ref{eq:uex_herl_1}). The continuity equation describes the interrelation between the gutter flow and the outer driving flows and it delivers the relation between the total flow ($Q_0$) and the volume flow fractions of the disperse ($Q_d$) as well as the continuous phase ($Q_c$). Considering the unit cell of a single slug and an adjoining droplet, the continuous phase parts into the volume flow of the slug ($Q_s$), through the gutters ($Q_g$) and through the wall film ($Q_f$):

\begin{equation}\label{eq:uex_herl_1}
Q_{0} = Q_{d} + Q_{c} = Q_d + \left( Q_{s} + Q_{g} + Q_{f}\right)
\end{equation}
\par

\nomenclature[U]{$s$}{slug}

\begin{figure}[htb]
	\centering
	\subfloat[][]{\includegraphics[width=0.7\linewidth]{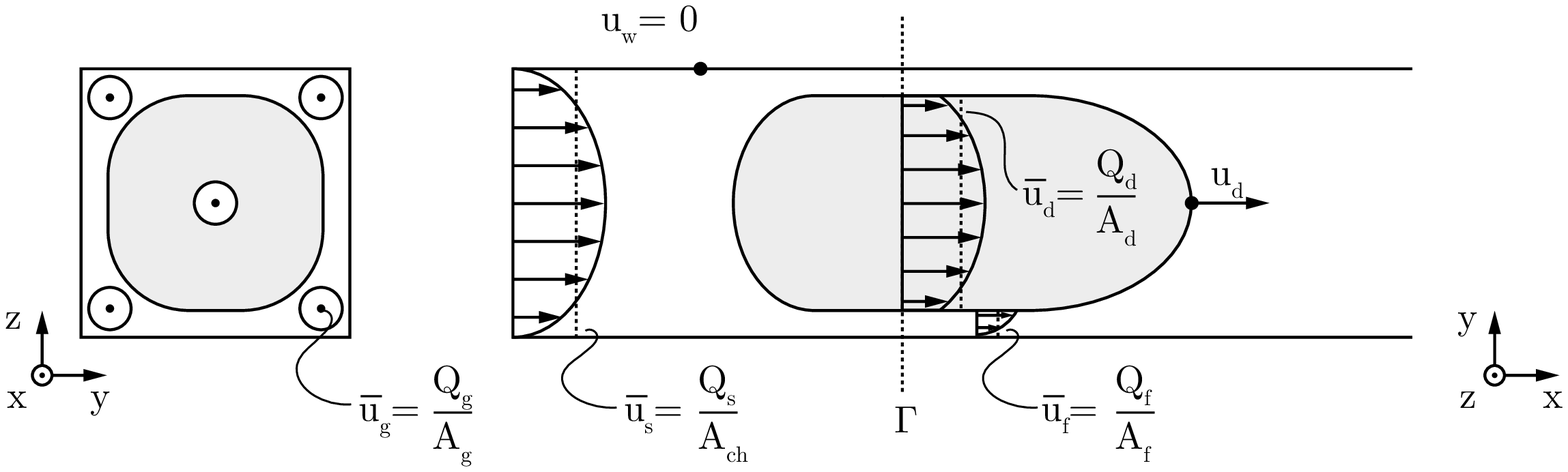}}%
	\qquad
	\subfloat[][]{\includegraphics[width=0.07\linewidth]{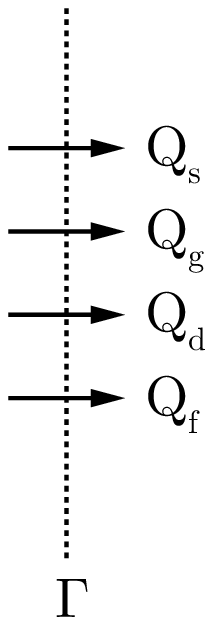}}%

	\subfloat[][]{\includegraphics[width=0.7\linewidth]{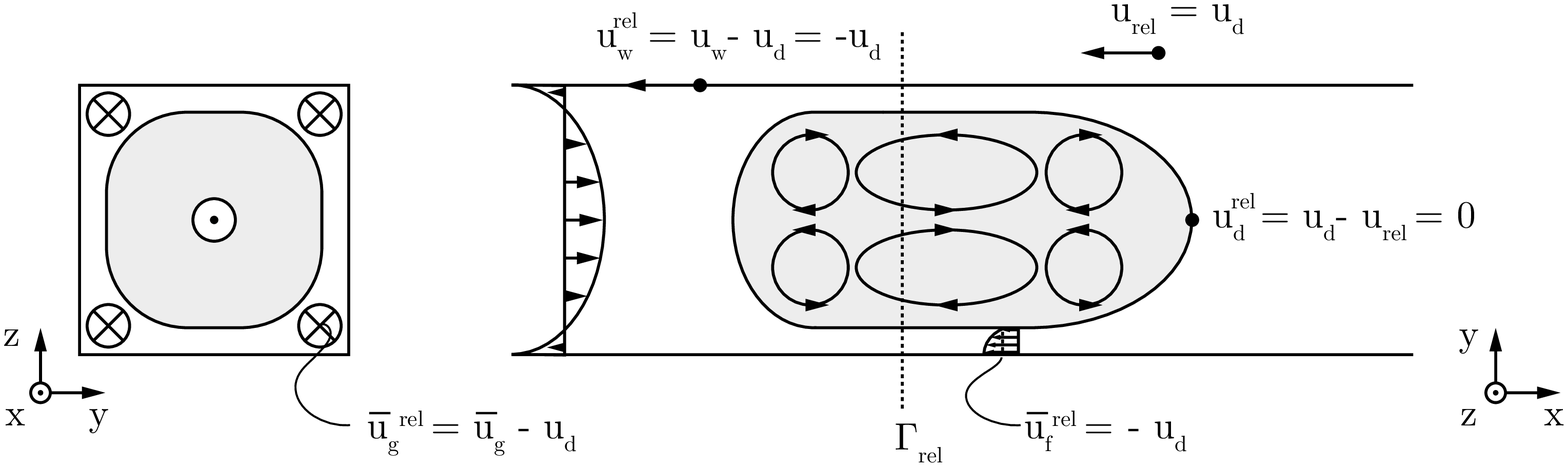}}%
	\qquad
	\subfloat[][]{\includegraphics[width=0.07\linewidth]{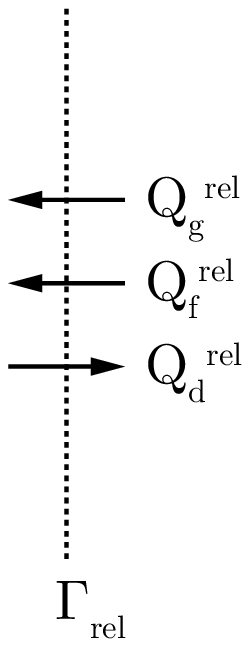}}%
\caption{Prominent averaged and local velocity for a flowing droplet, overlined entities represent area-averaged velocities. The film flow is not shown in this drawing. a) Velocities for a fixed point-of-view (Lagrangian-system), b) flow balance at a steady control surface c) velocities for a moving point-of-view with the velocity $u_d$ (Eulerian-system) d) flow balance at a moving control surface}
\label{fig:geschw_at_drop}       
\end{figure}

If we depict the flows at a moving Taylor droplet (Fig. \ref{fig:geschw_at_drop} a) ) and introduce a stationary control surface $\Gamma$ (Fig. \ref{fig:geschw_at_drop} b) ), velocities can be retrieved from the balances, while two differing flow states are possible: In the case of a droplet passing $\Gamma$, the slug volume flow through the control surface is $Q_s=0$ (since there is no slug present). For a slug passing $\Gamma$, only the slug volume $Q_s$ is present. As one can see, the use of a stationary point of view leads to instationary terms within the balances. 

If the coordinate system is changed from a Lagrangian to an Eulerian system by moving the control surface with an arbitrary velocity $u_{rel}$, the balances become stationary and relative velocities become visible (Fig. \ref{fig:geschw_at_drop} c) ). For this moving coordinate system, an additional volume flow $Q_{rel}$ adds to the balances, that results from the transformation of the coordinate system (Fig. \ref{fig:geschw_at_drop} d):

\begin{equation}\label{eq:uex_herl_1b}
Q_{rel} = u_{rel} \cdot A_{ch} 
\end{equation}
\par

This changes Eq. \ref{eq:uex_herl_1} to:

\begin{equation}\label{eq:uex_herl_2}
Q_{0} - Q_{rel} = Q_{d} + Q_{g} + Q_{f} - Q_{rel}
\end{equation}
\par

\nomenclature[U]{$rel$}{relative value}

\begin{figure}[htb]
\centering
\subfloat[][]{\includegraphics[width=0.6\linewidth]{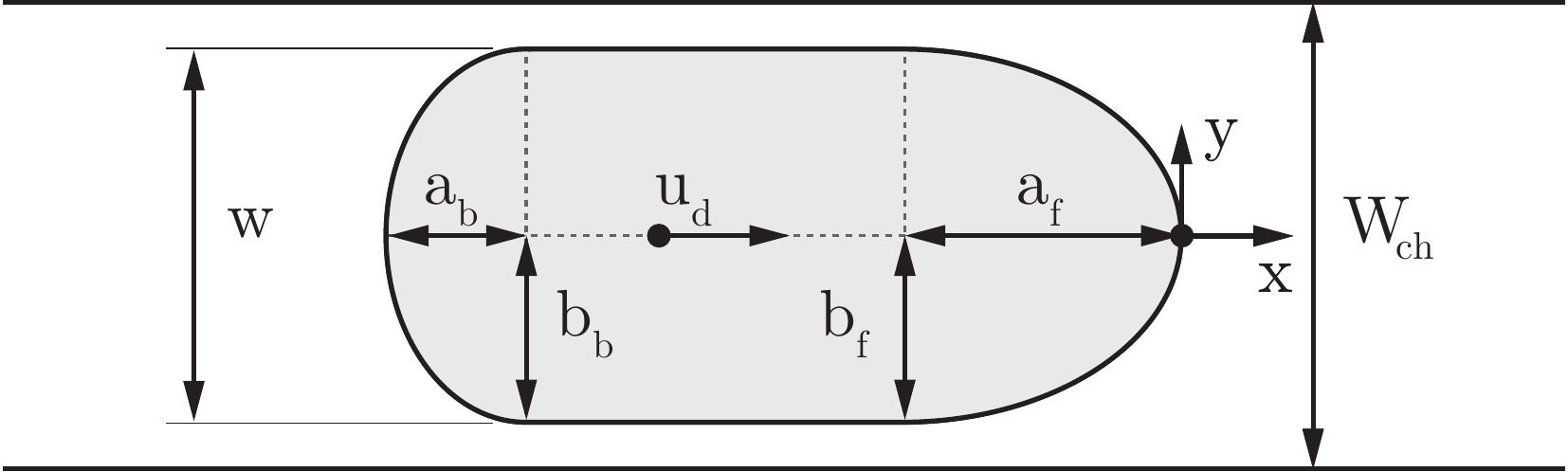}}%

\subfloat[][]{\includegraphics[width=0.25\linewidth, valign=c]{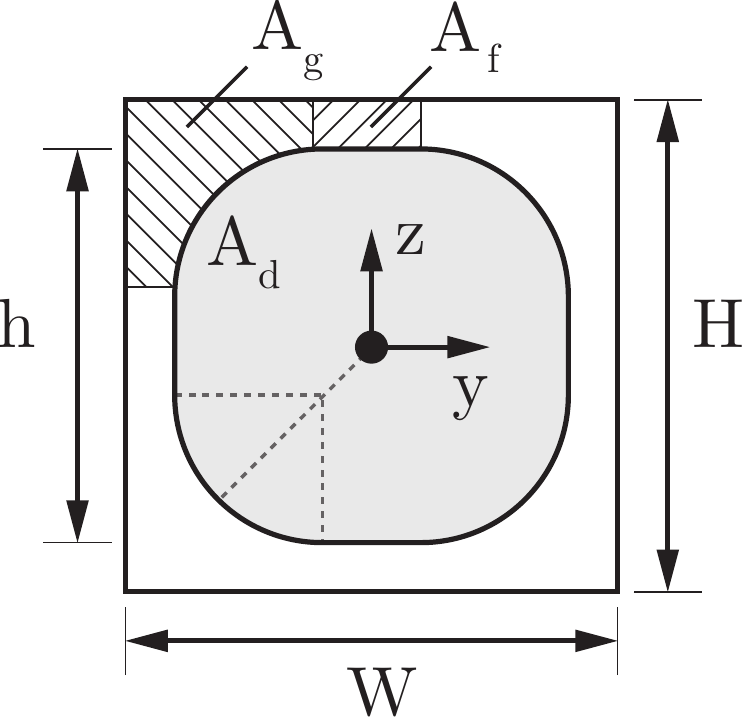}}%
\qquad
\subfloat[][]{\includegraphics[width=0.20\linewidth, valign=c]{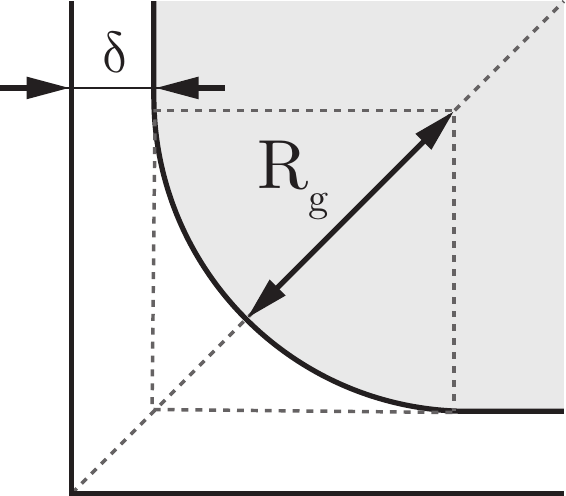}
\vphantom{\includegraphics[width=0.25\linewidth, valign=c]{masse_b.pdf}}}
\caption{Declaration of relevant geometry for the model of a Taylor-droplet flowing through a rectangular microchannel with the droplet velocity $u_d$. a) Top-view of x-y-plane, characterizing the droplet with the front ($a_f$, $b_f$) and back cap ($a_bf$, $b_b$), as well as the channel width $W$ and droplet width $w$ b) droplet front-view in y-z-plane with the droplet area $A_d$, gutter area $A_g$, film area $A_f$, channel height $H$ and the droplet height $h$. Only one representation of each area is shown c) close-up of the droplet corner region with gutter radius ($R_{g}$) and the film-thickness $\delta$}
\label{fig:groessen}       
\end{figure}

With knowledge of the specific areas for each distinct volume flow rate, the averaged velocities can be calculated. Following \ref{fig:groessen} a) + c)  we conclude for the gutter area $A_{g}$ of all four gutters 

\begin{align}\label{eq:Ag1}
A_{g} &=  4  \left[ \left( \overline{R_{g}}^2  - \frac{ \pi \overline{R_g}^2}{4} \right) + 2 \delta \overline{R_g} + \delta^2 \right] 
\end{align}
\par

with $\delta$ denoting the wall film thickness and $\overline{R_g}$ representing the mean dynamic gutter radius, which is described later in Sec. \ref{sec:ourmodel}. For the cross-sectional film area $A_{f}$ we state with the aspect ratio $ar = W H^{-1}$ 

\begin{align}\label{eq:Af1}
A_{f} &= \frac{4 \delta}{H} H^2 \left( \frac{1 + ar}{2} - 2 \frac{\overline{R_g} + \delta}{H}    \right)
\end{align}

The droplet cross-sectional area $A_d$ is delivered combining ${A_g}$ and $A_{f}$:

\begin{align}\label{eq:Ad}
A_{d} &= A_{ch} - \left( A_{g} + A_{f} \right)
\end{align}
\par

With the cross-sectional areas of the droplet, the gutter and the wall film, the volume flow rates can be rearranged to area-averaged velocities and Eq. \ref{eq:uex_herl_2} becomes

\begin{equation}\label{eq:uex_herl_3}
u_{0} A_{ch} - u_{rel} A_{ch} = \overline{u_{d}} A_{d} +  \overline{u_{g}} A_{g} +  \overline{u_{f}} A_{f} - u_{rel} A_{g} - u_{rel} A_{d} - u_{rel} A_{f}
\end{equation}
\par

\nomenclature[G]{$\Gamma$}{Control surface [-]}
\nomenclature[U]{$g$}{gutter}
\nomenclature[U]{$f$}{wall film}
\nomenclature[X]{$\overline{...}$}{area averaged value}

Herein $\overline{u_{d}}$, $\overline{u_{g}}$ and $\overline{u_{f}}$ are area-averaged velocities of the droplet, gutter and film. 

For the droplet ($Q_d = \overline{u_{d}} A_{d} - u_{rel} A_{d}$) and the film volume flow ($Q_f = \overline{u_{f}} A_{f} - u_{rel} A_{f}$), the transition velocity $u_{rel}$ equals the stagnation point velocity $u_{d}$, since we assume incompressibility, mass conservation, a stagnant film and a stationary droplet shape \citep{LuisA.M.Rocha.2017}

\begin{equation}\label{eq:uex_herl_3b}
\overline{u_{d}} =  u_{d} \overset{!}{=} u_{rel}\\
\end{equation}
\par

Additionally we assume the averaged velocity in the thin wall films to be insignificant for $Ca<0.2$

\begin{equation}\label{eq:uex_herl_4b}
\overline{u_{f}} \approx 0
\end{equation}
\par

Therefore Eq. \ref{eq:uex_herl_3} simplifies to

\begin{equation}\label{eq:uex_herl_7}
(u_{0}-u_{d}) A_{ch}  =  \overline{u_{g}^{rel}} A_{g} + (0 - u_d) A_f
\end{equation}
\par

Herein $\overline{u_{g}^{rel}} = \overline{u_{g}} - u_d$ represents the relative gutter velocity, which dissipates flow energy within the gutter. Simplification and combination with Eq. \ref{eq:ex} leads to 

\begin{equation}\label{eq:uex_herl_10}
\frac{u_{d}}{u_{0}} - 1  =  u_{ex} = - \frac{ \overline{u_{g}^{rel}} A_{g} }{ u_{0}   A_{ch}} + (u_{ex}+1) \frac{A_f}{A_{ch}}
\end{equation}
\par

\nomenclature[S]{$rel$}{relative}

We neglect the terms, that are small of higher order (see Appendix \ref{sec:uex_cons}) and retrieve an expression for the excess velocity:

\begin{equation}\label{eq:uex_herl_11b}
u_{ex} = -\frac{Q_g^{rel}}{Q_0} + \frac{A_f}{A_{ch}}
\end{equation}
\par

The relative gutter volume flow ($Q_g^{rel}$) and the cross-sectional area of the wall film ($A_f$) are the most prominent influencing quantities for the excess velocity. Thus our proposed model aims to especially determine these quantities.

\section{Model Specification}\label{sec:ourmodel}
The considerations of the previous section identify the volume flows significantly determining the excess velocity. In a second step, we clarify the relevant influential parameters on these volume flows and their interconnection. 

For the proposed modeling approach, we adapt a greybox model following \citet{Hangos.2001}. Our model is developed from engineering principles, hydrodynamic considerations (see Sec. \ref{sec:concept_exc}) and well-defined equations, whereas the initialization of a part of the influential parameters is based on measured data. The underlying relations can be described as an intermediate concept between a black box (completely based on measurement data) and a white box model (based only on analytically well-known equations and engineering principles).

As depicted in the previous chapter (Sec. \ref{sec:concept_exc}), we assume a droplet flowing through a rectangular microchannel with its properties: The thin wall film cross-section $A_f$ is determined by Eq. \ref{eq:Af1} and depends on the channel height $H$, the channel aspect ratio $ar$ and the film thickness $\delta$. To determine $\delta$, we apply the model of \citet{Han.2009}, which holds for $Ca<0.2$.

The relative volume flow through the gutter ($Q_{g}^{rel}$) is derived from the pressure difference along the gutters as suggested by \citet{Abiev.2017b}. Therefore knowledge of the relevant pressures is crucial. It can be determined from the dynamic interface deformation caused by the moving liquids through the gutters in flow direction \citep{Abate.2012}. Stagnant droplets have a static cap shape with a circular outline according to \citet{Musterd.2015}. When set into motion, the moving liquids exert forces onto the interface and cause a dynamic shape deformation \citep{LuisA.M.Rocha.2017}.

A model proposed by \citet{Miessner.2019} allows to approximate the droplet shape and gutter diameter. The model implies that the deformation difference of the dynamic droplet cap shape between the droplet front and back results in a change of the gutter radius from the static shape. The cross-sectional gutter area $A_g$ widens asymmetrically from back to front with the growing gutter radii to accommodate the relative volume flow $Q^{rel}_g$ of the continuous phase through the gutter. The gutter entrance at the droplet front is therefore larger than the gutter exit of the droplet back. Utilizing their model, the dimensionless radius of these gutters can be calculated from the flow related curvature of the droplet.

The gutter radius $k_{g,i}$ is therefore defined as a fraction of the droplet height $h$. For the case of present wall films the term is expressed as follows: 

\begin{equation}\label{eq:kg}
k_{g,i}=\frac{R_{g,i}}{h}=\frac{R_{g,i}}{H- 2 \delta}
\end{equation}
\par

In order to simplify the geometry, we define a mean gutter radius $\overline{R_{g}}$ using Eq. \ref{eq:kg}:

\begin{align}\label{eq:med_gut_rad2}
\overline{R_{g}} &=  \frac{k_{g,f}(H-2\delta)+k_{g,b}(H-2\delta)}{2}    \\
&= \frac{(H-2\delta)(k_{g,f}+k_{g,b})}{2}
\end{align}
\par

which is also used for the mean cross-sectional area $A_g$ derived with Eq. \ref{eq:Ag1}.

In previous work \citep{Helmers.2019}, we introduce a quantification of the droplet cap deformation with an elliptic approximation of the cap outline

\begin{equation}\label{eq:kfkf}
k_{c,f}=\frac{a_{f}}{b_{f}} 
\end{equation}
\par

\begin{equation}\label{eq:kfkb}
k_{c,b}=\frac{a_{b}}{b_{b}}
\end{equation}
\par

\nomenclature[U]{$c,f$}{front droplet cap}
\nomenclature[U]{$c,i$}{front or back droplet cap}
\nomenclature[U]{$c,b$}{back droplet cap}
\nomenclature[R]{$a$}{semi-minor axis of ellipse [$\mu m$]}
\nomenclature[R]{$b$}{semi-major axis of ellipse [$\mu m$]}

The ratio of the semi-major $a$ and semi-minor axis $b$ of the droplet cap curvature is introduced as deformation ratios $k_{c,i}$ at the droplet front and back. They become $k_{c,f} = k_{c,b} = 1$ when describing the static circular droplet cap shape of the droplet front and back. For the dynamic cap shape, under the influence of the moving liquids, the droplet front appears elongated $k_{c,f} > 1$ and the back cap is compressed in flow direction $k_{c,f} < 1$. Hence we introduce a correlation to describe these relations: The cap curvature only depends on the $Ca$-number for moderate flows ($Re<5$): 

\begin{equation}\label{eq:kfit}
k_{c,i} = m_{cap} \cdot Ca^{c_{cap}} + n_{cap}
\end{equation}
\par

\nomenclature[R]{$m$}{fitting coefficient [$-$]}
\nomenclature[R]{$n$}{fitting coefficient [$-$]}
\nomenclature[R]{$c$}{fitting coefficient [$-$]}
\nomenclature[U]{$cap$}{concerning droplet cap curvature}

With the correlation approach from our recent publication and the model from \citet{Miessner.2019}, we are able to calculate the Laplace pressure at the gutters. We assume a linear connection between the gutter front and the back of the droplet body, since the curvature of the gutter in flow direction is negligibly small. In this case, the mean interface curvature at the gutter entrance and its exit depends on the gutter radii only and the flow induced Laplace pressure difference equals:

\begin{equation}\label{eq:lapl_press_f}
\Delta p_{g,f} = \sigma ( \frac{1}{R^{stat}_{g,f}} - \frac{1}{R_{g,f}} ) 
\end{equation}
\par

\begin{equation}\label{eq:lapl_press_b}
\Delta p_{g,b} = \sigma ( \frac{1}{R^{stat}_{g,b}} - \frac{1}{R_{g,b}} ) 
\end{equation}
\par

\nomenclature[R]{$k$}{dimensionless ratio [$-$]}
\nomenclature[G]{$\delta$}{wall film thickness [$\mu m$]}
\nomenclature[R]{$R$}{radius [$\mu m$]}
\nomenclature[R]{$h$}{droplet height [$\mu m$]}
\nomenclature[U]{$g,i$}{i-th droplet gutter}
\nomenclature[U]{$g,f$}{gutter at front of droplet}
\nomenclature[U]{$g,b$}{gutter at back of droplet}
\nomenclature[S]{$stat$}{stationary}

Those deformation related pressures at the droplet front and back provide a link to the driving pressure difference $\Delta p_{LP}$ along the gutter length in the flow direction.  Due to symmetry, the static terms cancel out: 

\begin{align}\label{eq:lapl_diff1}
\Delta p_{LP,fb} &= \Delta p_{g,f} - \Delta p_{g,b} \nonumber \\
                 &= \sigma ( \frac{1}{R_{g,f}} -\frac{1}{R_{g,b}})  
\end{align}
\par

Using the dimensionless expression from Eq. \ref{eq:kg} this results in

\begin{align}\label{eq:lapl_diff2}
\Delta p_{LP,fb} &= \sigma ( \frac{1}{k_{g,f}(H-2\delta)}-\frac{1}{k_{g,b}(H-2\delta)})  \\
 &= (\frac{\sigma}{H}) \frac{1}{(1-\frac{2\delta}{H})} ( \frac{k_{g,b}-k_{g,f}}{k_{g,f} \cdot k_{g,b}}) 
\end{align}
\par

for the flow induced pressure difference as a driving force.

The relative volume flow rate through the four gutters $Q^{rel}_g$ can be modeled as a laminar pressure driven flow (\citep{Bruus.2008}) and it is linked to this pressure difference with a hydrodynamic resistance $\Omega$:

\begin{equation}\label{eq:q_gut_1}
Q_{g}^{rel} = \frac{1}{\Omega} \Delta p_{LP,fb}
\end{equation}
\par

\nomenclature[G]{$\Omega$}{flow resistance [$Pa \cdot s \cdot m^{-3}$]}
\nomenclature[R]{$l$}{length [$\mu m$]}
\nomenclature[G]{$\beta$}{geometric coefficient of resistance  [$-$]}

The hydrodynamic resistance $\Omega$ is defined by \citet{Ransohoff.1988} and \citet{Shams.2018} as

\begin{equation}\label{eq:q_gut_2}
\frac{1}{\Omega} =  \frac{\overline{R_{g}}^{2}}{\beta} \frac{A_{g}}{\eta_{c} \overline{l_{g}}}
\end{equation}
\par

Besides the mean gutter length $\overline{l_{g}}$, herein $\beta$ is a dimensionless factor, that represents the geometrical obstructions of the gutter flow, as well as the viscous coupling of both flow phases. In accordance with the simulation results of \citet{Shams.2018}, we declare an influence from the viscosity ratio $\lambda$ of the flow phases  to take care of the viscous coupling effects:

\begin{equation}\label{eq:betafit}
\beta = m_{\beta}\cdot \lambda^{c_{\beta}} + n_{\beta}
\end{equation}
\par

\nomenclature[U]{$\beta$}{concerning resistance factor}

The mean gutter length $\overline{l_{g}}$ can be derived from the droplet length $l_d$, if the gutter distance $\Delta x_{g,i}$ from the caps is subtracted:

\begin{equation}\label{eq:l_gut_1}
{\overline{l_{g}}} =  l_{d} - \Delta x_{g,f} - \Delta x_{g,b}
\end{equation}
\par

The gutter distance from the front and back droplet tip $\Delta x_{g,i}$ was defined by \citet{Miessner.2019} as

\begin{equation}\label{eq:deltax}
\Delta x_{g,i} = k_{c,i} \frac{H}{2}  \left[ \left( ar - \frac{2\delta}{H} \right) + \left( 1 -\frac{2\delta}{H} \right)  \left( 1 - 2 k_{g,i} \right) \right]
\end{equation}
\par

The above stated considerations lead to our final description for the relative volume flow through the gutters

\begin{equation}\label{eq:q_gut_3}
Q_{g}^{rel} = \frac{\overline{R_{g}}^{2}}{\beta} \frac{A_{g}}{\eta_{c} \overline{l_{g}}} \Delta p_{LP,fb}
\end{equation}
\par

\nomenclature[R]{$x$}{cap length till gutter entrance [$\mu m$]}

Herein $\overline{l_{g}}$ and $\Delta p_{LP}$ depend on $\overline{R_{g}}$ as the preceding considerations show. Thus $\overline{R_{g}}$ has the most prominent influence feature besides $\beta$. Inserting Eq. \ref{eq:q_gut_3} and Eq. \ref{eq:betafit} in Eq. \ref{eq:uex_herl_11b} delivers

\begin{equation}\label{eq:uex999+2}
u_{ex}  - \frac{A_f}{A_{ch}} = \frac{1}{Ca} \frac{1}{\beta} \frac{\overline{A_{g}}}{A_{ch}} \frac{\overline{R_g}^2}{l_g H} \frac{1}{(1-\frac{2\delta}{H})} ( \frac{k_{g,b}-k_{g,f}}{k_{g,f} \cdot k_{g,b}})
\end{equation}

expanding with $\frac{W}{W}$ and inserting $l_{g}$ (Eq. \ref{eq:l_gut_1}) we receive our final expression for the excess velocity:

\begin{equation}\label{eq:uex999+3}
u_{ex}  =  \frac{1}{Ca} \frac{1}{\beta} \frac{W}{l_d} \frac{\overline{A_{g}}}{A_{ch}} \frac{\overline{R_g}^2}{\left( 1 - \frac{\Delta x_{g,f}}{l_d}-\frac{\Delta x_{g,b}}{l_d} \right) A_{ch}} \frac{1}{(1-\frac{2\delta}{H})} ( \frac{k_{g,b}-k_{g,f}}{k_{g,f} \cdot k_{g,b}}) + \frac{A_f}{A_{ch}}
\end{equation}

We point out, that our model is usable within the capillary regime $Ca < 0.02$ since the model for the wall film-thickness from \citet{Han.2009}, the analytic interface model from \citet{Miessner.2019} and the droplet curvature correlation from our recent work \citep{Helmers.2019} are valid in this range. At higher $Ca$ in the viscous regime, different flow conditions with thicker wall films \citep{Jose.2014} as well as a higher influence of $Re$ are reported \citep{Kreutzer.2005c}.


\section{Model Calibration}\label{sec:fit_app}

The final expression for the excess velocity (Eq. \ref{eq:uex999+3}) depends on accessible data like $W$, $ar$ and $l_d$ as well as on $\overline{R_g}$, $\beta$, $k_{g,f}$, $k_{g,b}$, $\Delta x_{g,f}$ and $\Delta x_{g,b}$, whereas the latter are also related to the gutter radii $\overline{R_g}$. As shown in the previous section, the parameters can be calculated from the droplet cap curvatures $k_{c,f}$ and $k_{c,b}$ and via a measurement based model calibration. The 6 parameters $m_{cap,f}$, $m_{cap,b}$, $n_{cap,f}$, $n_{cap,b}$, $c_{cap,f}$, $c_{cap,b}$ influencing the cap deformation at the droplet front and back and additionally the dimensionless resistance factor $\beta$ with $m_{\beta}, c_{\beta}$ and $n_{\beta}$ need to be adjusted. 

For model calibration, we use the data-set presented in \citep{Helmers.2019} in combination with supplementary measurements in the data range of low $Ca$ and redefine $n_{cap}$ within an interval around 1. Nevertheless, for $Ca \rightarrow 0$ it represents the point of minimal surface energy and equates a spherical shape. This approach allows to improve the convergence of solvers. Out of further apriori considerations ($\beta > 0$, $n_{c,f} \approx 1$), we additionally define the boundaries for the search space of the solver in Tab. \ref{tab:boundaries}. We allow the solver to adapt the correlation coefficient from our last work to the presented model, since correlations of measurement data unavoidably include measurement errors, that might bias the solver results. 

\begin{table}[htb]
\centering
                \caption{Boundaries of input values for optimization}
                \label{tab:boundaries}       
                \begin{tabular}{l|r|r}
                \hline\noalign{\smallskip}
                parameter & lower boundary & upper boundary\\
                \noalign{\smallskip}\hline\noalign{\smallskip}
                $m_{cap,f}$  & 1.00 & 9.90 \\
                $c_{cap,f}$  & 0.40 & 1.50 \\
                $n_{cap,f}$  & 1.00 & 1.005 \\
                $m_{cap,b}$  & -2.50 & -1.00  \\
                $c_{cap,b}$  & 0.30 & 0.75\\
                $n_{cap,b}$  & 0.995 & 1.00\\
                $m_{\beta}$  & 0.00 & 10.00 \\
                $c_{\beta}$  & 0.50 & 1.5 \\
                $n_{\beta}$  & 0.50 & 20 \\
                \noalign{\smallskip}\hline
                \end{tabular}
            \end{table}
            
Beneath the fixed boundaries of the search space, a hydrodynamic boundary condition is applied to improve the convergence of the used optimization algorithms. For a rising $Ca$, the difference between $k_{g,f}$ and $k_{g,b}$ must increase, because of the pressure difference between the droplet front and back increases with higher $Ca$ \citep{Abiev.2017b} and the droplet front elongates, which leads to a larger front gutter radius, while the droplet rear flattens out. This is expressed by the gutter radius increase for an increasing $Ca$:

\begin{equation}\label{eq:randb}
\frac{d k_{g,f}}{d Ca} > \frac{d k_{g,b}}{d Ca} 
\end{equation}
\par

The large number of influence parameters leads to a highly nonlinear optimization problem with numerous local minima. Thus most gradient-based algorithms are not suitable for this type of optimization problem, since they tend to converge to local optima. This would result in an enormous number of randomly initialized solver calls to cover the whole search space. Thus an optimization algorithm that is capable of global optimization e.g. stochastic and metaheuristic approaches are more favorable to cover the search space and solve the statistical part of the greybox model. 

The quality of a solver result (e.g. deviation between measured data and estimation) is quantified by the loss-function of the problem. For the model calibration, we define

\begin{equation}\label{eq:loss-func}
\mathcal{L} = \omega_1 \left( \sum \frac{| k_{c,b} - k_{c,b}^{\mathcal{M}} | }{k_{c,b}^{\mathcal{M}}} + \sum \frac{| k_{c,f} - k_{c,f}^{\mathcal{M}} | }{k_{c,f}^{\mathcal{M}}} \right) + \omega_2 \sum \frac{|u_{d} - u_{d}^{\mathcal{M}}|}{u_{d}^{\mathcal{M}}} + \omega_3 \sum \frac{|\psi - \psi^{\mathcal{M}}|}{\psi^{\mathcal{M}}}
\end{equation}
\par

\nomenclature[G]{$\omega$}{weight factor [$-$]}
\nomenclature[S]{$\mathcal{M}$}{model calibration data}
\nomenclature[S]{$mod$}{model results}
\nomenclature[S]{$meas$}{measurement data}

with the weights of the individual properties $\omega_1$, $\omega_2$, $\omega_3$ following Tab. \ref{tab:omegas}. Values with an upper index $\mathcal{M}$ denote values estimated by the model and values without an upper index represent the calibration data. The first two sums serve as calibration data-sets for the hydrodynamic flow properties, since they contain the flow related deformation and therefore the hydrodynamic influences. The differences of the velocity $u_d$ serves as a parameter for the actual droplet velocity and therefore the flow resistance $\beta$. This is necessary, since otherwise no representation for $\beta$ is available. 

\begin{table}
\centering
                \caption{Weight factors $\omega_i$ of the loss-function for GA optimization}
                \label{tab:omegas}       
                \begin{tabular}{l|r|r}
                \hline\noalign{\smallskip}
                $\omega_1$ & $\omega_2$ & $\omega_3$\\
                \noalign{\smallskip}\hline\noalign{\smallskip}
               3.0  & 5.0 & 3.7 \\
                \noalign{\smallskip}\hline
                \end{tabular}
            \end{table}

An additional factor $\psi$ is introduced to maintain the overall integrity of the model: The excess velocity depends on extrinsic measurable values such as the dimensionless quantities and geometrical properties as well as varying flow properties like the gutter length. Thus, it is appropriate to separate the measurable quantities from the model based quantities. In doing so, one can directly compare the quantities received from our correlation with measurement data adjusted for material and flow properties. The separation of those terms leads to the equilibrium function $\psi$ (Eq. \ref{eq:psi1} + \ref{eq:psi1b}). Herein Eq. \ref{eq:psi1} represents the data from our measurements and Eq. \ref{eq:psi1b} only data from our modeling assumptions and geometry. For a well-adjusted model, the measured data for $\psi$ (upper equation) should correspond with the modeled values for $\psi$ (lower equation). 

\begin{align}
\psi &= \left( u_{ex} - \frac{A_f}{A_{ch}} \right) Ca ~ \beta \frac{l_d}{W} \label{eq:psi1}\\
\psi^{\mathcal{M}} &=  \frac{\overline{A_{g}}}{A_{ch}} \frac{\overline{R_g}^2}{\left( 1 - \frac{\Delta x_{g,f}}{l_d}-\frac{\Delta x_{g,b}}{l_d} \right) A_{ch}} \frac{1}{(1-\frac{2\delta}{H})} ( \frac{k_{g,b}-k_{g,f}}{k_{g,f} \cdot k_{g,b}}) \label{eq:psi1b} 
\end{align}

\nomenclature[G]{$\psi$}{Dimensionless validation function $[-]$}


The calibration procedure is based on a Genetic Algorithm (GA). Within the GA, every possible solution is emulated as a genetic code of population individuals. During the optimization process, the different solutions (individuals) can be combined (mated) to generate mixed solutions (children) with a combined genome. The decision, which solutions are actually combined, is based on the error (fitness value of loss function) of the solution. Within a so-called ranking sampling, the best solutions combine stronger than weak solutions (survival of the fittest), which leads to an improvement of the over-all population over the generations. Like in natural populations, random mutations of the genome can improve the overall fitness of a population. Transferred to a solver this means, that the population is able to leave local optima, if mutated individuals (solutions of partly random parameters) have a higher fitness value and therefore significantly change the genome pool of the population.

For good optimization results it is necessary to emulate a sufficiently sized genome pool, thus a high number of emulated individuals is preferred. This in turn results in a massively increased calculation demand, because for each iteration step every single individual and the children must be evaluated \citep{Whitley.1994}. Additionally, the genetic algorithm is typically performed several times to identify local optima. 

\begin{table}[htb]
\centering
                \caption{Properties of used solver algorithms}
                \label{tab:genalgparam}       
                \begin{tabular}{l|r}
                \hline\noalign{\smallskip}
                Genetic algorithm & \\
                \noalign{\smallskip}\hline\noalign{\smallskip}
                Population size   & 200 individuals\\
                Creation function   & Random feasable population\\
                Scaling function  & Ranking \\
                Selection function & Stochastic uniform \\
                Mutation function & Adaptive feasible \\
                Crossover function & Scattered \\
                \noalign{\smallskip}\hline
               Pattern Search algorithm & \\
                \noalign{\smallskip}\hline\noalign{\smallskip}
                Search method & Latin Hypercube \\  
                Poll method & Complete poll \\
                \noalign{\smallskip}\hline
                \end{tabular}
            \end{table}
            
To decrease the amount of time-consuming iteration steps of the GA and thereby reduce overall calculation time, we utilize a three-step stochastic and gradient free approach:

In the first step, a random population at the feasible borders of the problem is generated for the Genetic Algorithm and a genetic optimization performed. The convergence point of the GA is initialized via Latin Hypercube Sampling processed by a following fast-converging Pattern Search Algorithm (PSA) \citep{Davey.2008}, that results in an improved minimum as the final convergence point. The properties of both algorithms are shown in Tab. \ref{tab:genalgparam}. The algorithm finally merges at the values shown at the end of this section (Tab. \ref{tab:FITparam}). The results are discussed in the following.

\begin{figure}[htb]
\centering
 \includegraphics[width=0.6\linewidth]{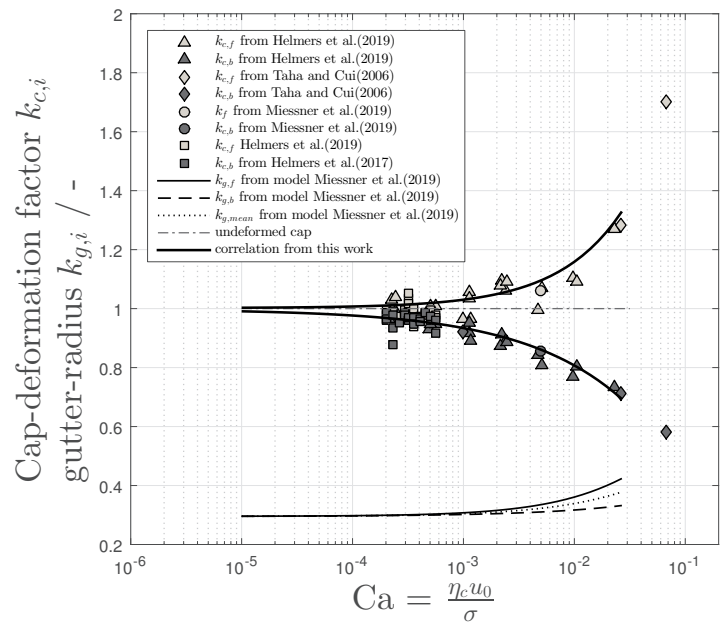}
\caption{Measured values and our model for the droplet cap deformation ratios $k_{c,f}$ and $k_{c,b}$ for different Ca-numbers. Additionally, the calculated dimensionless gutter-radii $k_{g,f}$, $k_{g,b}$ based on the model of \citet{Miessner.2019} are shown}
\label{fig:kfkb}       
\end{figure}

For the flow induced cap curvature, we again find the corresponding interrelation from our last work \citep{Helmers.2019}. For both cap deformation ratios $k_{f}$ and $k_{b}$ an exponential behavior as a function of the $Ca$-number is visible (Fig. \ref{fig:kfkb}). This proves the assumption of our previous work \citep{Helmers.2017} and agrees to \citet{Miessner.2019}. Our additional boundary condition (Eq. \ref{eq:randb}) is satisfied for all values as the graphs for the gutter radii $k_{g,i}$ show. 

Rising viscous forces, indicated by a rising $Ca$-number, deform the droplet interface more strongly. At static conditions ($Ca \rightarrow 0$) the cap curvatures are in correspondence with \citet{Musterd.2015} roughly circular ($k_{c,f}=k_{c,b} \approx 1$). The back cap is compressed with respect to the main flow direction ($k_{c,b} <1$) and the front cap elongated ($k_{c,f}>1$) if viscous forces rise in comparison to the static case.

The size of the gutter radii rises with the increasing influence of the viscous forces, since the bypass flow in the gutter increases and needs to be accommodated by the gutters. The gutter entrance is always larger in diameter than the exit radius ($k_{g,f} > k_{g,b}$).

\begin{figure}[htb]
\centering
 \includegraphics[width=0.6\linewidth]{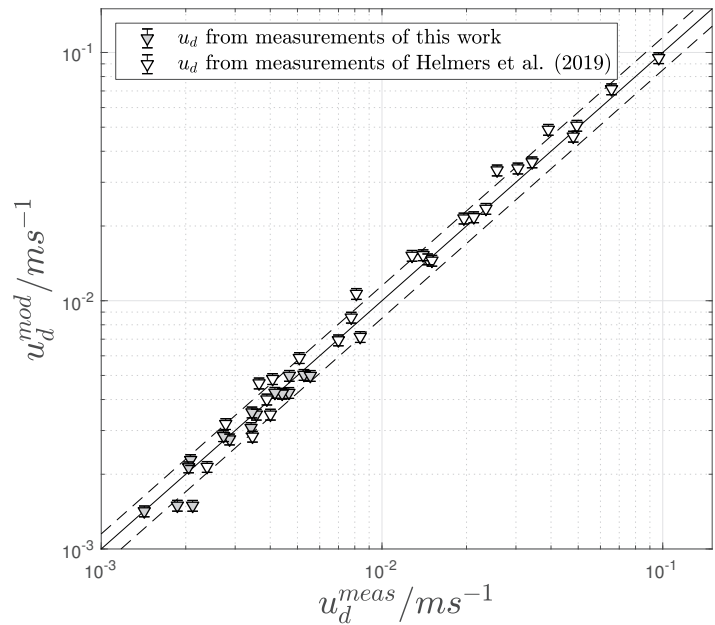}
\caption{Parity plot for the droplet velocity of measured droplets and corresponding data from our model after adaption of the dimensionless resistance factor $\beta$. The values situate fairly good within the fluctuation range of 10 \% as reported by \citet{Fuerstman.2007, VANSTEIJN.2008}}
\label{fig:parity}       
\end{figure}

For the resistance factor $\beta$ no fitting data is available since it can not be measured directly. Therefore $\beta$ is fitted based on the velocity data from high-speed camera measurements and the application of parameters for the droplet deformation. The resulting droplet velocities in comparison with the measured values are shown in Fig. \ref{fig:parity}. The measurements fit reasonably well within the range of inevitable velocity fluctuations caused by Taylor flow stability of +/- 10 \% range described by \citep{VANSTEIJN.2008}.

All adaption coefficients for the proposed model are summarized in Tab. \ref{tab:FITparam}.

\begin{table}[htb]
\centering
                \caption{Values for the fit-functions for droplet shape $k_{i}$ and resistance $\beta$}
                \label{tab:FITparam}       
                \begin{tabular}{l|r|r|r}
                \hline\noalign{\smallskip}
                Target value & $m_{i}$  & $c_{i}$  & $n_{i}$   \\
                \noalign{\smallskip}\hline\noalign{\smallskip}
                $ k_{f}$  &  4.8761 & 0.7465  & 1.0021 \\
                $ k_{b}$  & -1.6967 & 0.4745  & 0.9980 \\
                $ \beta$  &  6.1280 & 1.2105  & 1.4541 \\
                \noalign{\smallskip}\hline
                \end{tabular}
            \end{table}
\section{Model Validation}\label{sec:val}

\begin{figure}[htb]
\centering
 \includegraphics[width=0.6\linewidth]{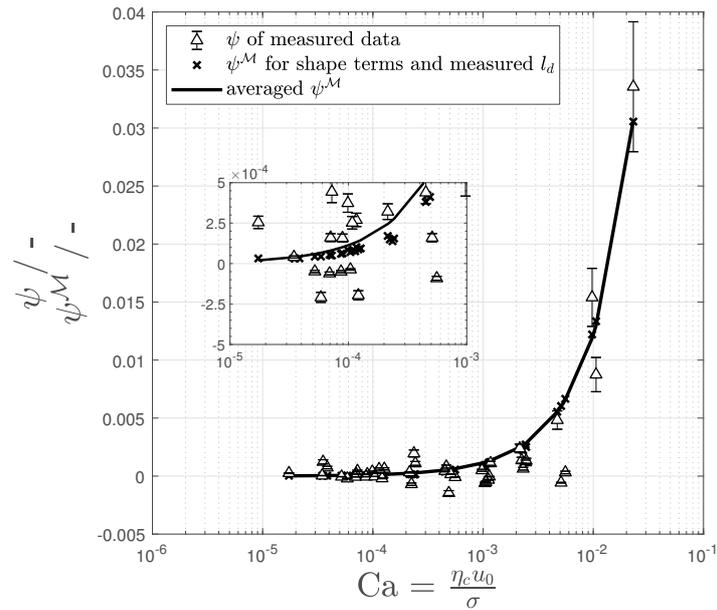}
\caption{Measured data $\psi$ (triangles) and modeled data $\psi^{\mathcal{M}}$ for averaged droplet lengths (solid line) over the Ca-number. Crosses depict the values for the actually measured droplet lengths}
\label{fig:fvonx}       
\end{figure}

A a first validation step the calibration functions $\psi$ and $\psi^{\mathcal{M}}$ are considered. For a hydrodynamic well-adjusted model, both function should coincide and our model based shape deviations ($\psi^{\mathcal{M}}$) equals the combination of measured properties ($\psi$). The corresponding data and the values for our model agree well at high $Ca$ numbers (Fig. \ref{fig:fvonx}), whereas a deviation for lower $Ca$-numbers can be observed. This can be explained by the fact, that the excess velocity itself is a relative quantity and it is therefore stronger influenced at lower absolute values (low $Ca$-number). Thus an inevitable constant measurement deviation for velocity and volume flows caused by the experimental equipment results in a higher error for low excess velocities. Especially for $Ca < 10^{-4}$ the resulting volume flows are situated at $Q_{0} \approx 2 \cdot 10^{-6} l/min$ and even minor deviations lead to high errors for the excess velocity. Thus, we consider our model approach to represent the measurements reasonably well and our fit coefficients to be valid.

\begin{figure}[htb]
\centering
 \includegraphics[width=0.6\linewidth]{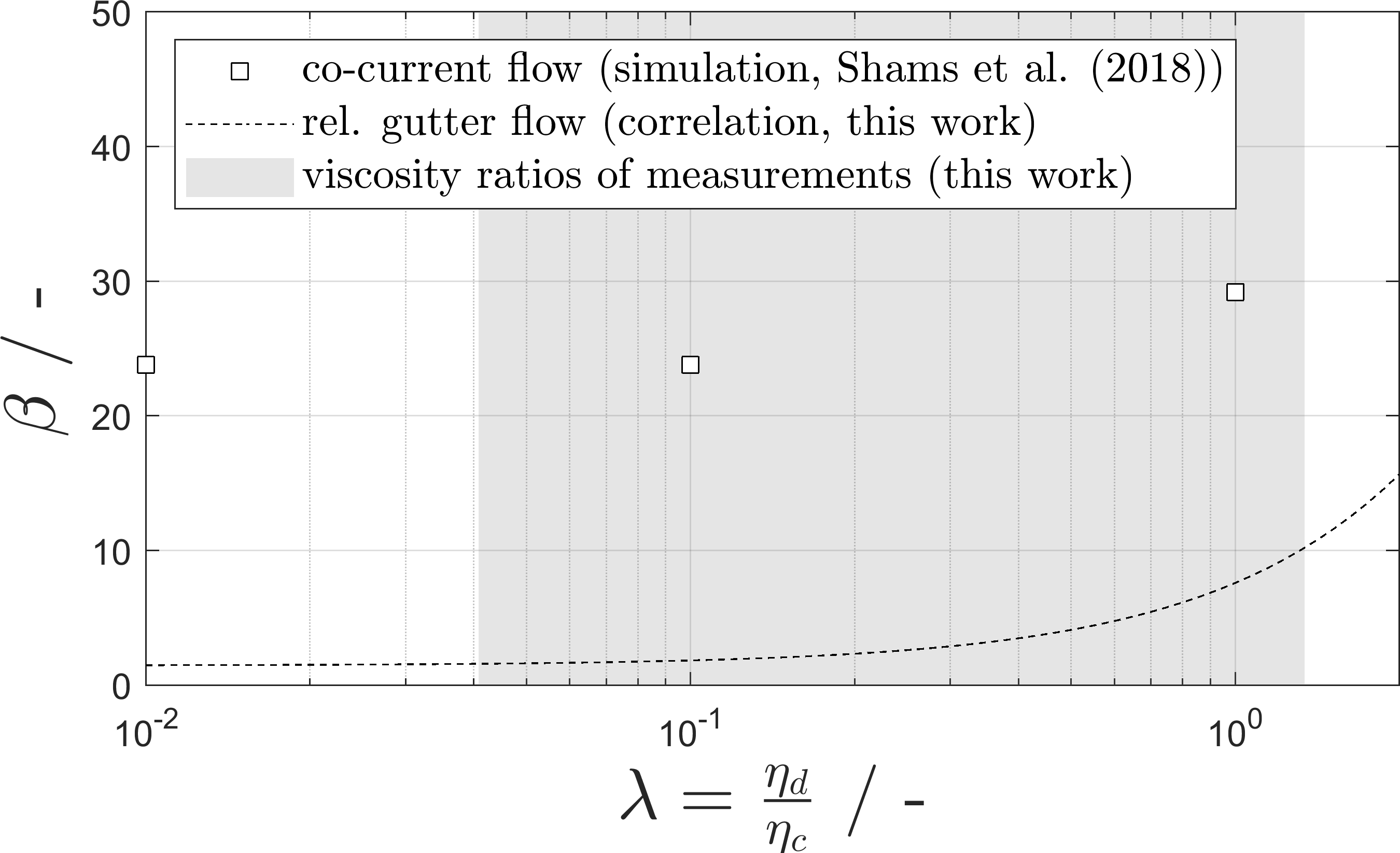}
\caption{Comparison of calculated resistance factor $\beta$ (squares) and correlation (dashed line) of our model. The data of the simulation from \citet{Shams.2018} is shown as squares. The $\lambda$-dependency of both works agree well, while our values are offset-shifted. This is caused by different flow specifications of both works: Eulerian specification \citep{Shams.2018}, Lagrangian specification (this work)}
\label{fig:shams_val}       
\end{figure}

Besides the hydrodynamic validation, our assumptions for $\beta$ are compared with available simulation data. We find a dependence of the gutter flow resistance $\beta$ on the viscosity ratio $\lambda$. The latter can be interpreted as an indicator for the viscous coupling of both flow phases (Fig. \ref{fig:shams_val}). In case of a highly viscous continuous phase ($\lambda<1$) the strongest velocity gradients are found inside the droplet, while for a viscous disperse phase ($\lambda>1$) larger velocity gradients and therefore energy dissipation is found inside the gutter-flow, resulting in a larger $\beta$.

This approach agrees with the simulations from \citet{Shams.2018}, who improved the model of \citet{Ransohoff.1988} by introducing the viscous coupling of the disperse and the continuous phase. For our case of $\lambda=0.1-1.4$ and a contact angle $\theta=0$ \citet{Shams.2018} report a $\beta$ between 20-30 for a co-current flow. Our values are shifted by a constant offset, while the slope and therefore the dependence on $\lambda$ is very similar. We consider this to be caused by the use of a different flow field specifications. \citet{Shams.2018} describe a concurrent flow in Eulerian specification, while in this work we determine $Q_g^{rel}$ within a Lagrangian flow specification. The coordinate transformation thus can only change the offset of the function, while the hydrodynamic influence (the slope) must remain identical. Additionally within the simulation of \citep{Shams.2018}, they assume a contact line between disperse phase, continuous phase and the wall in their problem definition. Although for the solution shown in Fig. \ref{fig:shams_val} the contact angle for the continuous phase is nearly 180 \textdegree, the existence of a contact line introduces an additional resistance. Thus regarding $\beta$ we consider our model valid.

\begin{figure}[htb]
\centering
 \includegraphics[width=0.96\linewidth]{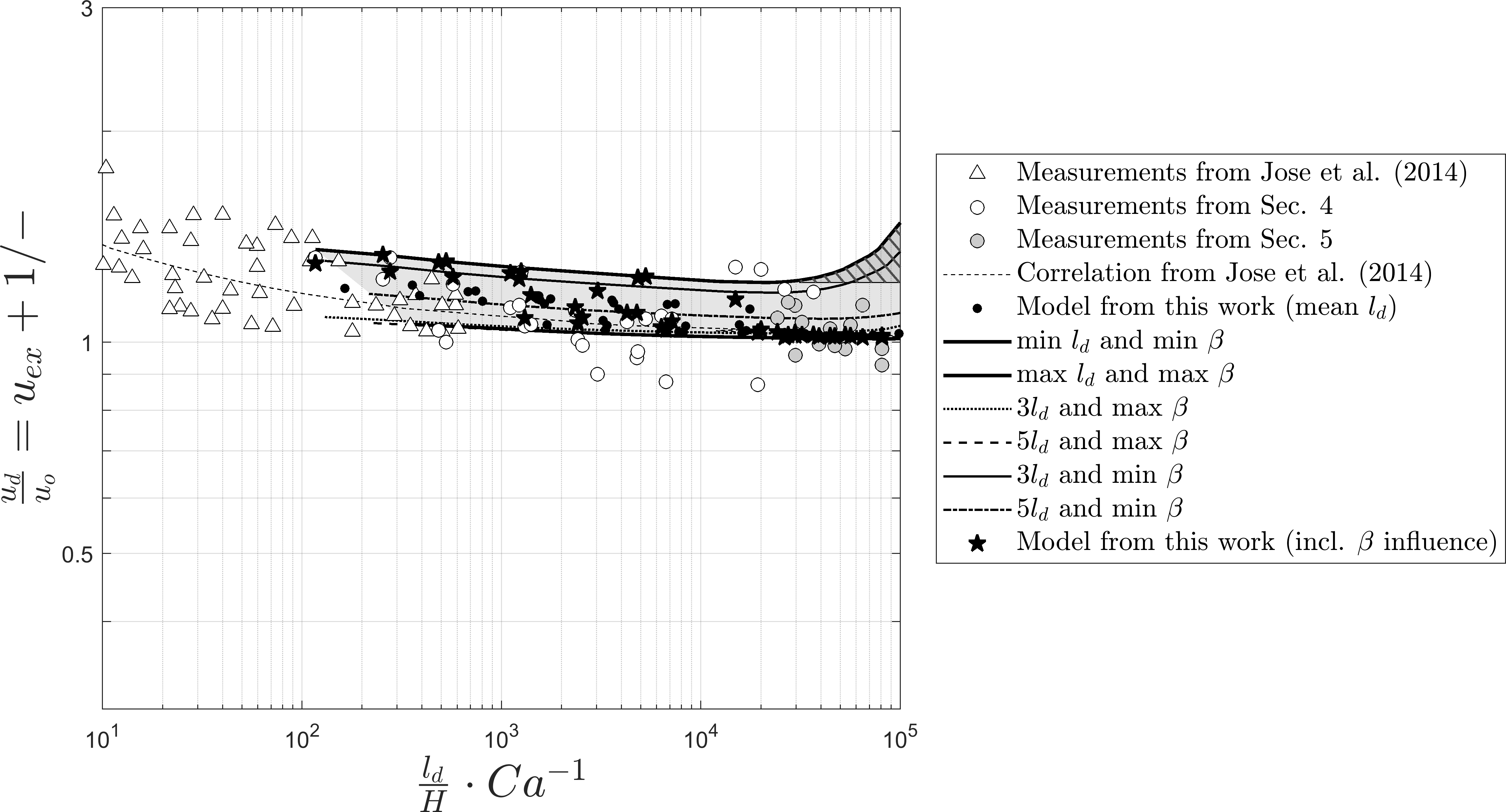}
\caption{Comparison of our model (stars) and measurements (triangles) with the measurements (circles) and correlation from \citet{Jose.2014} for a Taylor droplet in co-flow. The inclination for low $Ca$-numbers (hatched area) is discussed within the text. Since the influence of $l_d$ correlates linearly with $\Delta p_{LP}$ instead of $Ca$ in our model and $\beta$ is not included in the x-axis normalization, we additionally show the borders of our model for the minimum/maximum $l_d$ nad $\beta$ of our measurements}
\label{fig:cub_val}       
\end{figure}

The integrity of the model itself and the correlation for $\beta$ has been successfully proven, but due to the mentioned experimental restrictions, we can not directly compare the modeled and experimental determined excess velocities $u_{ex}$. Instead we compare the results of our model to different published approaches.

A suitable correlation for the prediction of the excess velocity in the first regime was introduced by \citet{Jose.2014}, who identify the $Ca$-number and the ratio of the droplet length $\frac{l_d}{H}$ as characteristic properties (Fig. \ref{fig:cub_val}). It has to be mentioned, that the model of \citet{Jose.2014} for non-wetting droplets ends at $\frac{l_d}{H Ca^{-1}} > 600$ , since for larger values they observe the disperse phase to wet the channel walls. This results in intensified dissipation and a higher pressure drop and thereby inhibits a gutter flow from the droplet front to the back. This phenomenologically equals the second flow regime as mentioned by \citet{WONG.2004}, but lacks the thin wall film and results in a much larger pressure drop. Furthermore, the viscosity ratio $\lambda$ is not included in their correlation.

A comparison of our deterministic model with \citet{Jose.2014} correlation Fig. \ref{fig:cub_val} shows good agreement. At very low $Ca$-numbers ($Ca < 10^{-4}$), our model results in a slightly increased excess velocity. We regard this behavior of the model as not physical. The effect results from the mathematical counterplay of the terms $\lim\limits_{Ca \rightarrow 0}{\frac{1}{Ca}} = \infty \leftrightarrow \lim\limits_{Ca \rightarrow 0} k_{g,f}-k_{g,b} = 0$ within Eq. \ref{eq:uex999+3}. In order to achieve a stable solver convergence, we accepted a small residual deviation for the static case $Ca \rightarrow 0$ for the front and back shape of $0.1 \% $. Due to the relative character of the excess velocity, this unfolds a significant influence at low $Ca$ values. Unfortunately, additional measurement validation concerning the shape deviation at very low $Ca$ is not possible in our experimental design, since the expected shape deviation is smaller than the blurriness of the interfacial area in the images itself and therefore lies within the measurement deviation.

\section{Discussion}
The interfacial area of a Taylor droplet in rectangular channels can be divided into the front and back cap regions, the wall films and the gutter interface. Neglecting the caps, the main momentum input into a droplet is transferred across the wall film and the gutter interface area. An increasing channel aspect ratio and droplet length result a growing wall film area, i.e. an enlarged dissipation interface. 

As we showed in Sec. \ref{sec:fluid_bas}, the behavior of the Taylor droplet's excess velocity can be parted in two possible regimes and the viscosity ratio $\lambda$ has a strong influence on the hydrodynamic mechanisms (Sec. \ref{sec:fluid_bas}). 

Within the first regime the fluid in the gutter flows slower than the droplet, exerting a drag force and leading to a positive excess velocity $u_{ex}>0$. These drag forces influence the droplet shape, leading to a flattened droplet back and elongated droplet front. We characterize this shape variation with a correlation (Sec. \ref{sec:fit_app}). As it can be seen from the measurements, our data falls into this flow regime, as our resulting droplet shape indicates in accordance with \citet{Wong.1995}.

The influence of the droplet length in the plug flow regime and $\lambda<1$, where larger droplets have an $u_{ex} \approx 0$ is in agreement with \citet{Jakiela.2011} and their later publication \citet{Jakiela.2012}. Additionally, they find an elongation of the dynamic droplet length in comparison to the static droplet length, which is also covered by our shape correlation, since for rising the $Ca$-number the droplet front elongates stronger than the droplet back is compressed. Recent published simulations by \citet{Kumari.2019} show, that also for larger $Re$-numbers $u_{ex} \approx 0$.

Our concept of deriving the excess velocity from the gutter pressure drop, that is inverted to the flow direction (larger pressure at the front gutter entrance) is in accordance with \citet{Abiev.2017b}. Nevertheless, the averaged pressure is still higher at the droplet back than on the droplet front due to the overall droplet pressure drop, since the droplet needs a driving force for its translation. 

For the second flow regime identified by \citet{Wong.1995}, where the viscous dissipation in the film and droplet leads to a bypass flow from the droplet back to the front and therefore a negative droplet excess velocity (for large $\lambda$ and long droplets), our model can be adapted, if the gutter-shape-difference term is revised or a resistance coefficient for the film is added. As the viscosity ratio $\lambda$ rises, more momentum will be dissipated via the wall films. This extra momentum is dissipated at the gutter interface, which results in a slower droplet velocity, forcing the continuous phase to bypass the droplet reversely. Since we are unable to establish a water-in-oil two-phase flow for a $\lambda \ll 1$ in our experimental setup due to the hydrophilized channel walls, droplet shape correlations for the case of $\lambda \gg 1$ should be performed in future work. Therefore, although we assume a systematic inversion of the gutter radii ratio back to front as a consequence of the reversed gutter flow direction, we like to mention that with the current shape correlation our model only works for the case of low viscous disperse phase ($\lambda \lessapprox 1$) like gas/liquid or low viscous oil/water flows.

The influence of the channel aspect ratio as mentioned by \citet{WONG.2004} is incorporated in our model: a higher aspect ratio results in lower excess velocities. This can be explained by the larger drag forces acting on a larger relative wall film area caused by the flattened channel geometry.

The comparison with the most prominent approaches shows, that our model and the chosen  influential parameters are valid for moderate and small viscosity ratios. The excess velocity is determined by viscous dissipation within the droplet and the gutters, as well as the drag of the thin wall films. The relation is characterized by the $Ca$-number, viscosity-ratio $\lambda$, the dimensionless gutter-length $l_g$, the aspect ratio $ar$ and the wall-film thickness $A_f$. Furthermore, the proposed model can close the gap for $\frac{l_d}{H Ca^{-1}} > 600$ and allows the calculation of the excess velocity for moderate $Ca$-numbers ($Ca < 0.02$).

\section{Conclusion}\label{10}
In this work, we developed a model to determine the relative droplet velocity of Taylor flows in square microchannels for moderate $Ca$-numbers and low to moderate viscosity disperse phase ($\lambda \lessapprox 1$). We base our model on the relative volume flow through the gutters, as well as the wall film thickness. The flow through the gutters is determined from the pressure drop described by the Laplace-Pressure difference between the gutter entrances. 

Our model uses the gutter radii to obtain the resulting pressure gradient that drives the continuous phase through the gutters. We use measurements at different $Ca$ and $Re$ in surfactant-free fluid system from a previous publication \citep{Helmers.2019} to derive the radii at the gutter entrances from the surface shape model proposed by \citet{Miessner.2019} and calibrate the model parameters. 

Our model is successfully validated with an intrinsic approach comparing the congruence of measurement data and calibrated model parameters. Additionally, we successfully compared our model to the phenomenological correlation of \citet{Jose.2014}.

For the future, our model should also be validated for aspect ratio differing from unity, since the influence of aspect ratio has been integrated into our model, but has not been validated so far. Additionally, the influence of surfactants and highly viscous droplets ($\lambda \gg 1$) on the excess velocity should be investigated to extend the model, since an excess velocity $u_{ex}<1$ was not included into the model so far. We suggest to include this function in the modelling of the wall film resistance. Especially local spatially resolved measurement techniques, e.g. \textmu -PIV measurements should be appropriate for this task.

\vspace{6pt} 



\authorcontributions{conceptualization, T.H. and U.M.; methodology, T.H. and U.M.; software, T.H., P.K. and U.M.; validation, T.H.; formal analysis, T.H. and U.M.; investigation, T.H.; resources, T.H.; data curation, T.H.; writing—original draft preparation, T.H.; writing—review and editing, U.M. and J.T.; visualization, T.H. and P.K.; supervision, U.M. and J.T.}

\funding{This research received no external funding.}


\conflictsofinterest{The authors declare no conflict of interest.} 

\abbreviations{The following abbreviations are used in this manuscript:
\printnomenclature[5em]
}

\appendixtitles{yes} 
\appendixsections{one} 
\appendix
\section{Considerations for $u_{ex}$} \label{sec:uex_cons}
Rearranging Eq. \ref{eq:uex_herl_10} leads to the equation 

\begin{equation}\label{eq:uex_herl_11}
u_{ex} - u_{ex} \frac{A_f}{A_{ch}}  = \frac{Q_g^{rel}}{Q_0} + \frac{A_f}{A_{ch}}
\end{equation}
\par

\begin{figure}[htb]
0
 \includegraphics[width=0.6\linewidth]{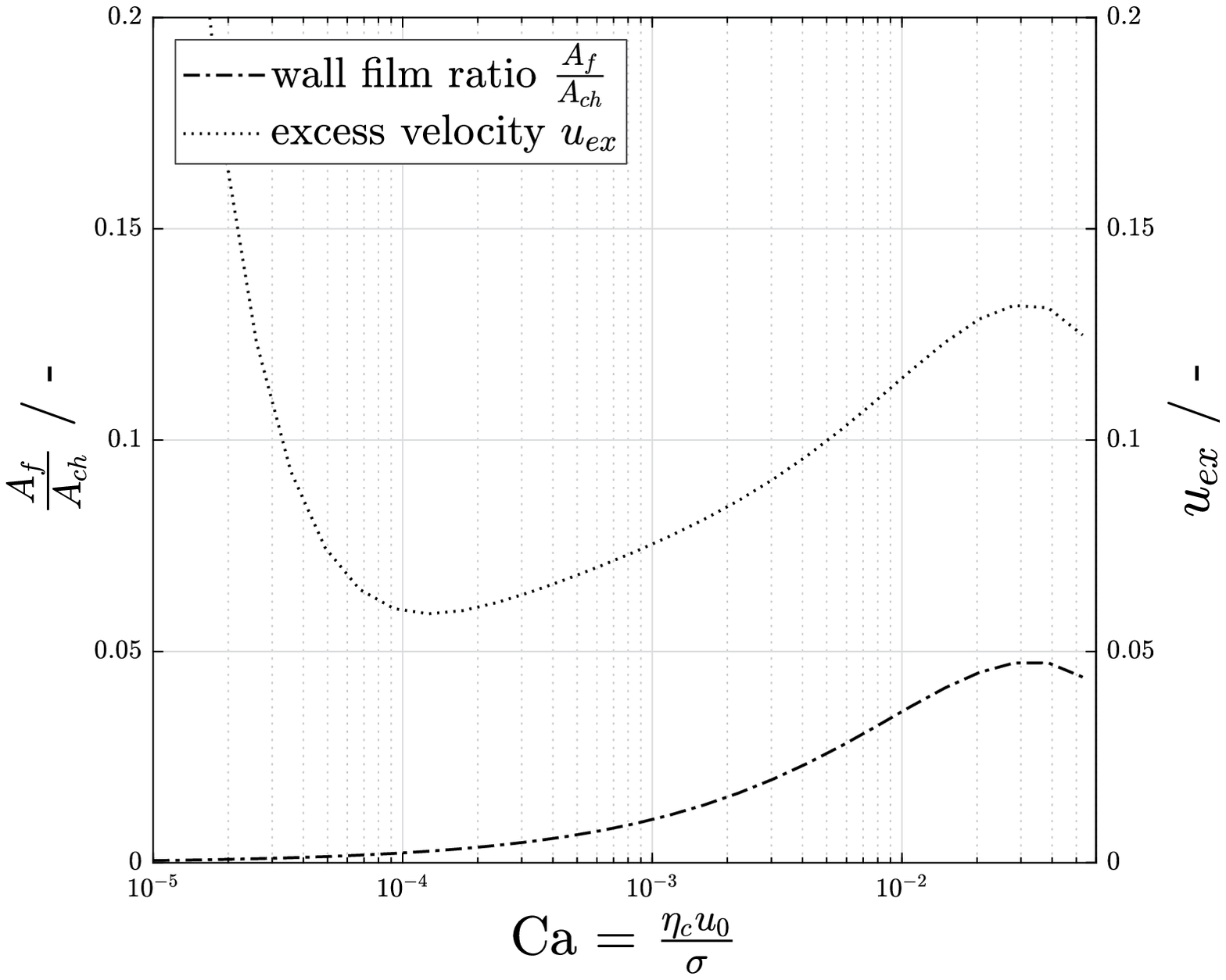}
\caption{Dimensionless film-area and excess velocity for $\frac{l_d}{W} = 3$ and $\beta = 3.892$ for the proposed model}
\label{fig:mischterm}       
\end{figure}

Our measurements show in agreement with \citet{Jose.2014} excess velocities with values $u_{ex}<0.4$ for $Ca<0.2$. Additionally we can assume $\frac{A_f}{A_{ch}}<0.005$ as shown in Fig. \ref{fig:mischterm}. Thus one can say $u_{ex} \frac{A_f}{A_{ch}} < 0.002$ and therefore it can be considered small of higher order and be neglected.

\unskip
\externalbibliography{yes}
\bibliography{references.bib}



\end{document}